\documentclass{article}

\PassOptionsToPackage{numbers, compress}{natbib}

\usepackage[final]{neurips_2023}

\usepackage[utf8]{inputenc} % allow utf-8 input
\usepackage[T1]{fontenc}    % use 8-bit T1 fonts
\usepackage[colorlinks=true,linkcolor=cyan,urlcolor=red]{hyperref}       % hyperlinks
\usepackage{url}            % simple URL typesetting
\usepackage{booktabs}       % professional-quality tables
\usepackage{amsfonts}       % blackboard math symbols
\usepackage{nicefrac}       % compact symbols for 1/2, etc.
\usepackage{microtype}      % microtypography
\usepackage{xcolor}         % colors
\hypersetup{
    colorlinks=true,
    linkcolor=red,
    filecolor=magenta,
    citecolor=blue,
    urlcolor=blue,
    }
\usepackage{amsmath}
\usepackage{multirow}
\usepackage{subcaption}
\usepackage{graphicx}

\title{Adversarial Fine-tuning using Generated Respiratory Sound to Address Class Imbalance}
\author{%
  June-Woo Kim$^{1,4}$, Chihyeon Yoon$^{1}$, Miika Toikkanen$^{2}$, Sangmin Bae$^{3,4}$, Ho-Young Jung$^{1}$\thanks{corresponding author}\\
  $^1$Department of AI, Kyungpook National University\\ 
  $^2$ALI Co., Ltd, $^3$KAIST AI, $^4$RSC LAB, MODULABS\\
  Republic of Korea\\
  \texttt{\{kaen2891, chichi8969, hoyjung\}@knu.ac.kr}\\
  \texttt{miika.toikkanen.2@gmail.com}, \texttt{bsmn0223@kaist.ac.kr}
}

\begin{document}

\maketitle

\begin{abstract}
%\jw{The scarcity of medical data is a major obstacle to the development of high-performing medical AI systems.} 
%Deep generative models have emerged as a promising approach to address this challenge, and have achieved impressive results in the medical image domain. However, generating satisfactory samples of medical sequential data, such as respiratory sound has received less attention.
Deep generative models have emerged as a promising approach in the medical image domain to address data scarcity. However, their use for sequential data like respiratory sounds is less explored.
%generating satisfactory samples of sequential data, such as respiratory sound, remains a challenging problem. 
In this work, we propose a straightforward approach to augment imbalanced respiratory sound data using an audio diffusion model as a conditional neural vocoder. 
We also demonstrate a simple yet effective adversarial fine-tuning method to align features between the synthetic and real respiratory sound samples to improve respiratory sound classification performance.
%In this work, we propose a straightforward approach to augment imbalanced respiratory sound data using an audio diffusion model as a conditional neural vocoder. We further introduce a simple yet effective adversarial adaptation method to align the synthetic and original respiratory sound samples for improving respiratory sound classification performance.
Our experimental results on the ICBHI dataset demonstrate that the proposed adversarial fine-tuning is effective, while only using the conventional augmentation method shows performance degradation.
%Our experimental results on the ICBHI dataset with generated samples demonstrate that the proposed adversarial adaptation is effective in aligning synthetic and original data, while using only conventional augmentation method shows performance degradation.
Moreover, our method outperforms the baseline by 2.24\% on the ICBHI Score and improves the accuracy of the minority classes up to 26.58\%. For the supplementary material, we provide the code at \url{https://github.com/kaen2891/adversarial_fine-tuning_using_generated_respiratory_sound}.
 %\jw{Moreover, our method outperforms the baseline by 2.24\% on the ICBHI Score and considerably improves the accuracy of the imbalanced abnormal class by 26.58\%.}
%Moreover, our method achieves an improvement of 3.76\% ICBHI Score over the baseline and considerably improves the performance on the imbalanced abnormal class by \jw{N\%}.
%Moreover, our method achieves an improvement of 3.76\% ICBHI Score over the baseline and significantly improves the performance on the imbalanced abnormal class by \jw{N\%}.
%It is better to avoid the word "significant" unless we have done a statistical test. Consider instead: considerably, notably, clearly
%The code will be available in the final version.

%Despite the notable advances of generative models in deep

\end{abstract}

% 흘림체: \emph{}
% Bold: \textbf{}
% 강조? \verb+

%keywpoint word: High-fidelity samples
\section{Introduction}
%\jw{Recent deep learning technologies have shown remarkable advances and achieved impressive performance in several practical applications. However, the issues with scarcity of data in the medical domain limit the effectiveness of deep learning models and pose a significant challenge for training medical AI models. Such problems could lead to the risk of failures in critical scenarios and call for finding solutions to better utilize the limited data.}
Deep generative models (DGMs) have become popular due to their potential to address data scarcity via augmentation. Among recent advancements, methods such as generative adversarial networks (GANs)~\cite{goodfellow2014generative}, variational autoencoders (VAEs)~\cite{kingma2013auto}, and diffusion probabilistic models ~\cite{ho2020denoising} are gaining attraction.
Notably, previous studies~\cite{antoniou2017data, bowles2018gan} have demonstrated that training a model on a mixture of real samples and synthetic samples generated by DGMs can be an effective approach to better utilize the limited data. %addressing the challenge of data scarcity. 
In medical domain, DGMs have been successfully leveraged to synthesize medical data in a variety of categories, including retinal images~\cite{costa2017end, iqbal2018generative}, CT and MRI scans~\cite{nie2017medical, shin2018medical, sandfort2019data}, as well as X-rays~\cite{loey2020within, motamed2021data}.
%In medical domain, %Furthermore, 
%DGMs have been successfully leveraged to synthesize medical data in a variety of categories, including retinal images~\cite{costa2017end, iqbal2018generative}, CT and MRI scans~\cite{nie2017medical, shin2018medical, bowles2018gan, sandfort2019data, han2019synthesizing, armanious2020medgan, wolleb2022diffusion, dorjsembe2022threedimensional}, as well as X-rays~\cite{loey2020within, motamed2021data}. 
%, which received additional interest due their applicability in diagnosing COVID-19 cases.
%\jw{However, synthesis is more challenging when it comes to sequential medical data, including respiratory sound~\cite{kochetov2020generative, jayalakshmy2021conditional, saldanha2022data}, EEG recording~\cite{tian2023dual}, and ECG signals~\cite{narvaez2020synthesis} due to its complex temporal dynamics, high dimensionality and relative lack of benchmarks.}
However, synthesis is more challenging when it comes to sequential medical data, including respiratory sound~\cite{kochetov2020generative, jayalakshmy2021conditional, saldanha2022data}, due to its complex temporal dynamics, high dimensionality and relative lack of benchmarks. %, EEG recording~\cite{tian2023dual}, and ECG signals~\cite{narvaez2020synthesis} due to its complex temporal dynamics, high dimensionality and relative lack of benchmarks.
%However, the generated samples are generally less satisfactory when it comes to sequential medical data, including respiratory sound~\cite{kochetov2020generative, jayalakshmy2021conditional, saldanha2022data}, \jw{EEG recording~\cite{tian2023dual}, and ECG signals~\cite{narvaez2020synthesis}.}
%This is because audio data involves intricate temporal dynamics, higher dimensionality, more challenging evaluation metrics, and a lack of comprehensive datasets compared to image data.

\begin{figure}
    \centering
    \includegraphics[width=0.92\linewidth]{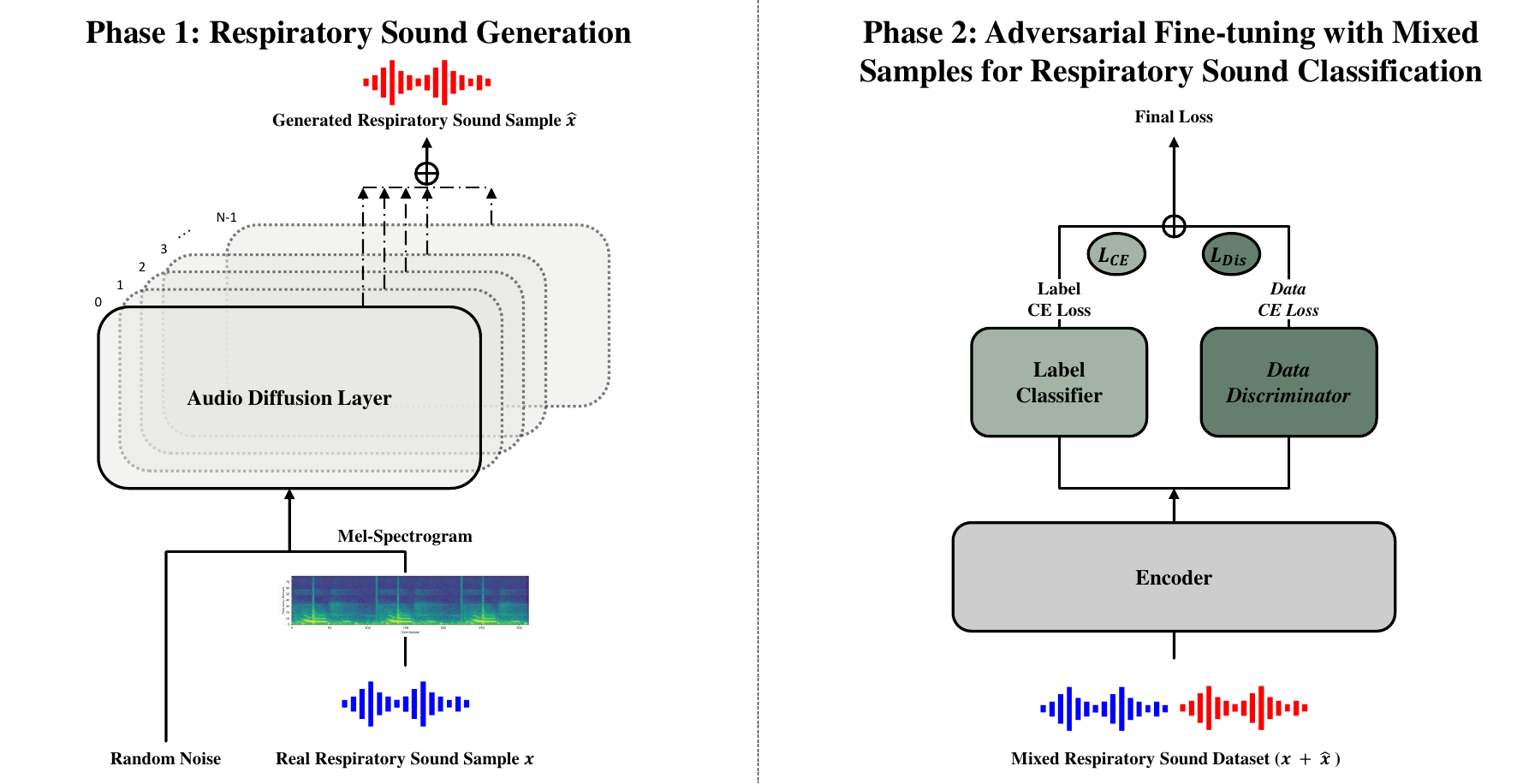} 
    \caption{In phase 1, we generate the respiratory sound samples using the audio diffusion model as conditional neural vocoder. In phase 2, we use the proposed adversarial fine-tuning method to address the distribution inconsistency between synthetic and real samples for training the respiratory sound classification model.}
    \label{fig:overview}
\end{figure}

%In this paper, we first aim to generate high-fidelity respiratory sound samples using DGMs. 
%In this paper, we aim to generate high-fidelity respiratory sound samples using DGMs and then combine these synthetic samples with real data to improve the performance of respiratory sound classification models, especially for imbalanced lung sound disease classes.
In this paper, we aim to generate high-fidelity respiratory sound samples using DGMs and then combine these synthetic samples with real data to improve the respiratory sound classification task, especially for imbalanced lung sound disease classes.
%We then combine and align these synthetic samples with real data to improve the performance of respiratory sound classification models, especially for imbalanced lung sound disease classes. 
Figure~\ref{fig:overview} illustrates the overall process of our approach split into phases 1 and 2.
%illustrates the overall process of our approach. 
In phase 1, we introduce a simple method for generating respiratory sound samples using a conditional neural vocoder inspired by the recent success of audio diffusion models~\cite{ho2020denoising, kong2020diffwave} in obtaining realistic audio.
%However, the discrepancy between synthetic and real samples can introduce problems related to distribution inconsistency, which can degrade the performance of respiratory sound classification models as the proportion of synthetic data in the training set increases.
However, the discrepancy between synthetic and real samples can introduce problems related to distribution inconsistency,
%combining synthetic and real data can introduce distribution mismatch problems, 
which can degrade the performance of respiratory sound classification models as the proportion of synthetic data in the training set increases.
In phase 2, we propose a simple yet effective \emph{adversarial fine-tuning} method motivated by~\cite{ganin2016domain} to learn the model that is distribution-agnostic between real and synthetic data.
%To address the issue of distribution mismatch, we propose a simple yet effective \emph{adversarial adaptation} method motivated by~\cite{ganin2016domain} to learn model that is distribution-agnostic between real and synthetic data.
%to align the synthetic and original respiratory sound samples. 
%In phase 2 in Figure~\ref{fig:overview}, the proposed adversarial adaptation works by training a discriminator to distinguish between synthetic and real samples, while simultaneously training a classifier to predict the respiratory sound label.
The adversarial fine-tuning method relies on a discriminator network feedback to move features obtained from real and synthetic samples closer to each other, while simultaneously training a classifier to predict the respiratory sound label.
%This training process forces the classifier to learn features that are invariant to the data distribution mismatch.%, allowing the model to learn these features regardless of whether the samples are synthetic or real.

Our experimental results on the ICBHI~\cite{rocha2018alpha} dataset demonstrate that the proposed adversarial fine-tuning method effectively aligns the features from synthetic and real data, leading to improved performance while simply combining synthetic and real samples for training resulted in performance degradation. 
Specifically, our method achieves a 2.24\% ICBHI Score improvement over the baseline and up to 26.58\% accuracy improvement of the minority classes.
%\jw{Specifically, our method achieves a 2.24\% improvement over the baseline and a considerably 26.58\% accuracy improvement on the imbalanced abnormal class.}
%Specifically, our method achieves a 3.76\% improvement over the baseline and a considerably \jw{5\%} improvement on the imbalanced abnormal class. 
Our contributions are: $(i)$ We show the successful generation of high-fidelity respiratory sound samples with audio diffusion model as conditional neural vocoder $(ii)$ We demonstrate adversarial fine-tuning on respiratory sound data, %We introduce adversarial adaptation, 
which can overcome data distribution inconsistency between synthetic and real samples $(iii)$ We present that the proposed method enables the synthetic and real training samples to be used more effectively, considerably improving performance in the imbalanced abnormal lung disease class.

%Checking citation~\cite{bae23b_interspeech}

%However, the experimental results with mixed samples with real and generated respiratory sounds show that the more data amounts, the lower the performance.

%\section{Related work}

\section{Method}
\paragraph{Audio Diffusion Probabilistic Model} 
Diffusion probabilistic models~\cite{ho2020denoising} are a type of deep generative model that use a Markov chain to gradually add Gaussian noise $\mathcal{N}(x_{t};\sqrt{1-\beta_{t}}x_{t-1},\beta_{t}I)$ into a complex data distribution. 
The posterior $q(x_{1},...,x_{T}|x_{0})$ called \emph{diffusion process} or \emph{forward process} is defined by a fixed Markov chain that transforms the input data $x_{0}$ to a latent variable $x_{1},...,x_{T}$ according to a variance schedule $\beta_{1},...,\beta_{T}$:
\begin{equation}
    q_{\theta}(x_{1},...,x_{T}|x_{0})\! := \prod_{t=1}^{T}q(x_{t}|x_{t-1}),\, \, \, \, \, \, \, \, \, \, \, \,    q(x_{t}|x_{t-1}):=\mathcal{N}(x_{t};\sqrt{1-\beta_{t}}x_{t-1},\beta_{t}I)
\end{equation}
The joint distribution $p_{\theta}(x_{0},...,x_{T-1}|x_{T})$ called \emph{reverse process} is defined by a Markov chain with learned Gaussian transitions starting at $p(x_{T})=\mathcal{N}(x_{T}; 0, I)$:
\begin{equation}
    p_{\theta}(x_{0},...,x_{T-1}|x_{T})\! = \prod_{t=1}^{T}p_{\theta}(x_{t-1}|x_{t})
\end{equation}
where the transition probability $p_{\theta}(x_{t-1}|x_{t})$ is parameterized as $\mathcal{N}(x_{t-1};\mu_{\theta}(x_{t},t),\sigma_{\theta}(x_{t},t)^{2}I)$ with shared parameter $\theta$, and the both $\mu_{\theta}$ and $\sigma_{\theta}$ are calculated with the diffusion-step and $x_{t}$. 

In this work, we use the audio diffusion model~\cite{kong2020diffwave} which is neural vocoding conditioned on Mel-spectrogram as a conditional neural vocoder to reconstruct the respiratory sound raw waveform. To this end, we employ the \verb|DiffWave_BASE| model for our audio diffusion model, which consists of a stack of 30 residual layers, each with 64 residual channels as well as bidirectional dilated convolution with kernel size 3, and the dilation is doubled at each layer within each block. To ensure that the output of the audio diffusion model has the same length as the Mel-spectrogram, a transposed 2D convolution upsampler is provided for the conditioned 2D Mel-spectrogram. 

In our conditional neural vocoder setting, the outputs of the upsampler are added to the dilated convolutions in each residual layer for reconstruction. In other words, our audio diffusion model is conditioned on the Mel-spectrogram, which means that it uses a lot of prior knowledge to guide the generation process. This makes it easier to generate realistic samples and reduces the need for large amounts of training data. This is especially beneficial in the medical domain, where data is often scarce.

\paragraph{Adversarial Fine-tuning} While augmenting real data with synthetic samples can be beneficial, we found in early experiments that in our case the distribution mismatch between the two types of data degraded the performance of the classification model.
%Combining synthetic and real data for data augmentation can be beneficial, but the distribution mismatch between the two types of data can degrade the performance of the classification model as the proportion of synthetic data in the training set increases. 
To overcome this issue, we propose a simple yet effective \textbf{A}dversarial \textbf{F}ine-\textbf{T}uning (AFT) method inspired by~\cite{ganin2016domain}. The proposed method consists of two losses with label classifier $\mathcal{L}_\text{CE}$ and data discriminator $\mathcal{L}_\text{Dis}$:
\begin{align}\label{eq:sdat1} %tilde -> hat
\mathcal{L}_{\text{CE}} \! = -\frac{1}{N}\sum_{i=1}^n\! \, y_{i}\! \, \log \, \!(\hat{y_{i}}), \quad \mathcal{L}_{\text{Dis}} \! = -\frac{1}{N}\sum_{i=1}^n\! \, d_{i}\! \, \log \, \!(\hat{d_{i}}).
%\mathcal{L}_{\text{DAT}} = \mathcal{L}_{\text{CE}}^y + \lambda \, \cdot \! \mathcal{L}_{\text{CE}}^d
\end{align}
where $\mathcal{L}_{\text{CE}}$ and $\mathcal{L}_{\text{Dis}}$ are CE loss with label $y$ and data type label $d$, and the predicted probabilities $\hat{y}$ and $\hat{d}$ are obtained by passing through the classifier and data discriminator, respectively. To ensure that the learned features cannot distinguish between the synthetic and real samples, gradients from $\mathcal{L}_{\text{Dis}}$ are multiplied by a negative constant during the backpropagation. The final training objective is $\mathcal{L}_{\text{Final}} = \mathcal{L}_{\text{CE}} + \lambda \, \! \mathcal{L}_{\text{Dis}}$ where $\lambda$ is a regularization parameter drawn from~\cite{ganin2016domain}. The AFT aims to reduce classification error while ensuring learned features are consistent across data types.
%In other words, the goal of the adversarial adaptation method is to minimize the classification error while simultaneously encouraging that the learned features are invariants from the data type.
%We then show the proposed adversarial adaptation can overcome the data distribution mismatch between synthetic and real samples. 

\section{Experimental Setup}
\paragraph{ICBHI and Mixed Dataset} We used the ICBHI~\cite{rocha2018alpha} dataset for respiratory sound tasks, following the official train-test split (60/40\%). 
The training (4,142) and test sets (2,756) contain four classes: \emph{normal} (49.8\%/57.29\%), \emph{crackle} (29.3\%/23.55\%), \emph{wheeze} (12.1\%/13.97\%) and \emph{both} (8.8\%/5.19\%), hereinafter referred to as $C_{n}$, $C_{c}$, $C_{w}$ and $C_{b}$, respectively.
We generated synthetic samples for each minority class to balance them with the majority class, and then mixed these with real data. We denote it \emph{Mixed-ICBHI} datasets as follows: \emph{Mixed-500, ..., Mixed-N, ..., Mixed-5k} where the number $N$ refers to the total amount of samples per class. 
We prioritize real samples so that synthetic samples are only added if the sample count is less than $N$.
%We prioritize real samples so that synthetic samples are only added if the sample count is less than $N$ (i.e., \emph{Mixed-500} only contains synthetic samples from $C_{both}$). 
We used the \emph{Specificity}, \emph{Sensitivity} and their arithmetic mean, hereinafter referred to as $S_{p}$, $S_{e}$, and $Score$, respectively~\cite{rocha2018alpha}. For ICBHI details and additional statistics on the Mixed-ICBHI dataset, see Appendix~\ref{appendix_icbhi} and~\ref{appendix_mixed_dataset}.

\paragraph{Audio Diffusion Model}
For the data pre-processing, we fixed all of the data length as 4 seconds and extracted the 4,142 respiratory sound samples from the ICBHI dataset as 80-dimensional Mel-spectrograms. For the audio diffusion model, we trained the DiffWave~\cite{kochetov2020generative} on the ICBHI training set from scratch. To this end, we used a linearly spaced schedule for the diffusion variance schedule parameter $\beta_{t}$ $\in [1\times10^{-4}, 0.02]$, 50 and 6 denoising diffusion steps for training and evaluation, respectively. We then trained the model for 1M training steps with Adam~\cite{kingma2014adam} optimizer, a learning rate of 1e-4, and a batch size of 16. 

\paragraph{Respiratory Sound Classification} To prepare the data for training, we fixed the duration of all synthetic and real samples to 5 seconds and extracted 128-dimensional log Mel filterbank features with a window size of 25 ms and an overlap size of 10 ms. We then normalized the log Mel filterbank features using the mean and standard deviation of -4.27 and 4.57, as described in~\cite{bae23b_interspeech}. We trained the classification model using pretrained Audio Spectrogram Transformer~\cite{gong21b_interspeech} (AST) model with the Adam optimizer, a learning rate of 5e-5, and a batch size of 32 for 50 epochs. To ensure the stability of our results, we trained our model using a fixed set of five random seeds for all experiments.

\begin{table}[!t]
    \centering
    \caption{Respiratory sound classification performance on ICBHI test set according to various mixed sample amounts using the AST~\cite{gong21b_interspeech} fine-tuning as described in~\cite{bae23b_interspeech}. No Aug. denotes only the real ICBHI dataset is used for training. We only report the ICBHI \emph{Score} (\%). \textbf{Bold} denotes the best result.}\label{tab:table1}
    % \vspace{-0.2pt}
    % \vspace*{-5pt}
    \addtolength{\tabcolsep}{0pt}
    \resizebox{\linewidth}{!}{
    \begin{tabular}{l|cccccccc}
    \toprule
                                              & \multicolumn{7}{c}{training dataset}                                    \\
    \cmidrule(l{2pt}r{2pt}){2-9}
    method                                                                 & No Aug. & Mixed-500 & Mixed-800 & Mixed-1k & Mixed-1.5k & Mixed-2k & Mixed-3k & Mixed-5k \\
                                                                     \hline \midrule
                                                                     
    \begin{tabular}[c]{@{}l@{}}AST FT\end{tabular}                                                               & 59.55 & 59.92$_{\pm 0.82}$    & 59.99$_{\pm 1.17}$    & 59.81$_{\pm 0.36}$   & 59.65$_{\pm 0.30}$     & 59.18$_{\pm 0.65}$   & 59.04$_{\pm 0.32}$   & 58.56$_{\pm 0.84}$   \\
    \midrule
    \begin{tabular}[c]{@{}l@{}}AFT\end{tabular} & - &  \textbf{61.79}$_{\pm 0.47}$    & 60.89$_{\pm 0.78}$    & 60.8$_{\pm 1.05}$    & 60.03$_{\pm 1.14}$     & 60.64$_{\pm 0.45}$   & 59.96$_{\pm 0.38}$   & 59.74$_{\pm 0.6}$  \\
    \bottomrule
    \end{tabular}%

    }
\end{table}

\section{Results} 
\subsection{Effectiveness of Adversarial Fine-tuning}
To validate the proposed AFT, we compared it against AST fine-tuning (AST FT) with only cross-entropy (CE) loss on several Mixed-ICBHI datasets under the same conditions.
As in Table~\ref{tab:table1}, the AST FT performance decreased as the number of augmented samples in the ICBHI dataset increased, while the AFT outperformed it in each case, reaching the best Score on \emph{Mixed-500}.
Based on the result, as $N$ increases, the distribution mismatch between synthetic and real samples increases, therefore leading to reduced performance. Our method mitigates this to a degree, but still benefits more in smaller $N$. We further explore how our method affects the performance of minority classes. We report their accuracy on the ICBHI test set for AST FT with no augmentation, and AST FT and AFT on Mixed-500 and Mixed-2k.
As in Table~\ref{tab:table2}, directly fine-tuning on mixed data did not improve the performance of the minority classes overall. However, our proposed method improved their accuracies by up to 26.58\%, especially in $C_{b}$.
These results show that our method can most effectively enhance the performance of minority classes despite using synthetic samples that would otherwise degrade them.
For additional confusion matrices of Table~\ref{tab:table2}, see Appendix~\ref{appendix_confusion}.
%To explore how our method affects the performance of minority classes, we report their accuracy on the ICBHI test set for AST FT with no augmentation, and AST FT and AFT on Mixed-500 and Mixed-2k.
%To explore how our method affects the performance of minority classes, we report their accuracy on the ICBHI test set for the following trained models: AST FT with no augmentation, and both AST FT and AFT on Mixed-500 and Mixed-2k, respectively.
%As in Table~\ref{tab:table2}, directly using mixed data does not improve the performance of AST fine-tuning. However, our proposed method achieves a significant 26.58\% improvement in the performance of $C_{b}$. 

%To demonstrate how our method improves the performance of abnormal minority classes, we present the accuracy of these classes on the ICBHI test set, according to different amounts of Mixed-ICBHI dataset used for model training.
\subsection{Comparison on ICBHI Dataset Results}

\begin{table}[!t]
    \centering
    \caption{Accuracy (\%) of the abnormal class on the ICBHI test set for AST fine-tuning and AST adversarial fine-tuning models trained on different datasets. %Accuracy (\%) of the abnormal class on the ICBHI test set according to various amounts of Mixed-ICBHI dataset used to train the model. 
    \textbf{Bold} denotes the best result.}\label{tab:table2}
    % \vspace{-0.2pt}
    % \vspace*{-5pt}
    \addtolength{\tabcolsep}{2pt}
    \scriptsize{
    \resizebox{\linewidth}{!}{
    \begin{tabular}{lc|c|cc|cc}
    \toprule
     &  & \multicolumn{5}{c}{method (dataset)} \\
    \cmidrule(l{2pt}r{2pt}){3-7}
     & & AST FT & AST FT & AST FT & Adversarial FT & Adversarial FT\\
     class & ratio & (No aug.) & (Mixed-500) & (Mixed-2k) & (Mixed-500) & (Mixed-2k) \\
                                      \hline \midrule
    %Normal                            & 83.98        & 84.36         & 81.82        & 80.3  & 79.73 \\
    %\midrule
    crackle ($C_{c}$) & 23.55\%                          & 45.45        & 42.84         & 42.86         & 44.07  & \textbf{46.99} \\
    % \midrule
    wheeze ($C_{w}$) & 13.97\%                           & 36.62        & 36.1         & 22.08        & \textbf{37.92}  & 30.12   \\
    % \midrule
    both ($C_{b}$) & 5.19\%                              & 15.38         & 9.09         & 7.69        & \textbf{41.96} & 35.66 \\ 
    \bottomrule
    \end{tabular}
    }}
\end{table}

\begin{table*}[!t]
    \centering
    \caption{Overall comparison of the ICBHI dataset for the respiratory sound classification task. We compared previous studies that followed the official 60-40\% split for the training/test set. Scores marked with $*$ denote the previous state-of-the-art performance. \textbf{Best} and {\underline{second best}} results.}
    
    \label{tab:table3}
    \renewcommand{\arraystretch}{1}
    \addtolength{\tabcolsep}{3pt}
    \resizebox{\linewidth}{!}{
    \begin{tabular}{l|lc|lll}
    \toprule
    method & architecture & pretrain & $S_p$\,(\%) & $S_e$\,(\%) & \textbf{Score}\,(\%) \\
    \hline \midrule
    RespireNet \cite{gairola2021respirenet} (CBA+BRC+FT) & ResNet34 & IN  & 72.30 & 40.10  & 56.20 \\
    
    %\multirow{18.5}{*}{\rotatebox[origin=c]{90}{\textbf{4-class eval.}}} & SE+SA \cite{yang2020adventitious} & ResNet18 & - & \textit{INTERSPEECH`20} & {81.25} & 17.84 & 49.55 \\
    %& CNN-MoE \cite{pham2021cnn} & C-DNN & - & \textit{JBHI`21} & 72.40 & 21.50 & 47.00 \\
    %& RespireNet \cite{gairola2021respirenet} & ResNet34 & IN & \textit{EMBC`21} & 71.40 & 39.00 & 55.20 \\

    %& SE+SA \cite{yang2020adventitious} & ResNet18 & -  & {81.25} & 17.84 & 49.55 \\
    %& LungRN+NL \cite{ma2020lungrn2} & ResNet-NL & -  & 63.20 & 41.32 & 52.26 \\
    
    %& LungBRN \cite{ma2019lungbrn} & bi-ResNet & -  & 69.20 & 31.12 & 50.16 \\
    %& CNN-MoE \cite{pham2021cnn} & C-DNN & -  & 72.40 & 21.50 & 47.00 \\
    %& Ren \textit{et al.} \cite{ren2022prototype} & CNN8-Pt & -  & 72.96 & 27.78 & 50.37 \\
    Wang \textit{et al.} \cite{wang2022domain} (Splice) & ResNeSt & IN  & 70.40 & 40.20 & 55.30 \\
    % & Nguyen \textit{et al.} \cite{nguyen2022lung} & ResNet50 & IN & \textit{TBME`22} & 76.33 & 37.37 & 56.85 \\
    %Nguyen \textit{et al.} \cite{nguyen2022lung}\,(StochNorm) & ResNet50 & IN & 78.86 & 36.40 & 57.63 \\
    Nguyen \textit{et al.} \cite{nguyen2022lung}\,(CoTuning) & ResNet50 & IN  & 79.34 & 37.24 & $\text{58.29}$ \\
    %& Moummad \textit{et al.} \cite{moummad2022supervised} & CNN6 & AS & \textit{arXiv`22} & 70.09 & 40.39 & 55.24 \\
    Moummad \textit{et al.} \cite{moummad2022supervised}\,(SCL) & CNN6 & AS  & 75.95 & 39.15 & 57.55 \\    
    Bae \textit{et al.} \cite{bae23b_interspeech}\, (Fine-tuning)  & AST & IN\,+\,AS  & $\text{77.14}$ & $\text{41.97}$ & $\text{59.55}$ \\
    %& Bae \textit{et al.} \cite{bae23b_interspeech}\, (Patch-Mix CE)  & AST & IN\,+\,AS & \textit{INTERSPEECH`23} & $\text{76.87}$ & $\text{42.04}$ & $\text{59.46}$ \\
    Bae \textit{et al.} \cite{bae23b_interspeech}\, (Patch-Mix CL) & AST & IN\,+\,AS  & $\text{\bf {81.66}}$ & $\text{\bf{43.07}}$ & $\text{\bf 62.37}^\textbf{*}$ \\
    \cmidrule{1-6}
    \textbf{AFT on Mixed-500 [ours]} & AST & IN\,+\,AS  & $\text{\underline{80.72}}_{\pm 0.99}$ & $\text{\underline{42.86}}_{\pm 1.3}$ & $\text{\underline{61.79}}_{\pm 0.47}$ \\
    \bottomrule
    \end{tabular}}
\end{table*}

Table~\ref{tab:table3} presents an overall comparison of various methods for lung sound classification on the ICBHI dataset. %, including our proposed method. 
Our proposed method trained with Mixed-500 achieved a Score of 61.79\%, outperforming the AST FT model by 2.24\%, which is comparable to the state-of-the-art model.
%Our proposed method trained with Mixed500 achieved a Score of 61.79\%, outperforming the AST fine-tuning model by 2.24\%. 
This demonstrates the efficacy and potential of our proposed method, indicating its capability for addressing the issues with synthetic data.
%Consequently, our method can align features between synthetic and real samples, thus effectively addressing data distribution inconsistency.
%We would like to emphasize that our results notably improved the performance for imbalanced abnormal lung disease classes.
%\jw{We would like to emphasize that our method can address the class imbalance problem, and can be helpful in detecting disease in the real world.}
%We would like to emphasize that our method can align features between synthetic and real samples, therefore effectively addressing data distribution inconsistency.

\subsection{Qualitative Analysis}
\begin{figure}[!ht]
    \centering
    \includegraphics[width=1\linewidth]{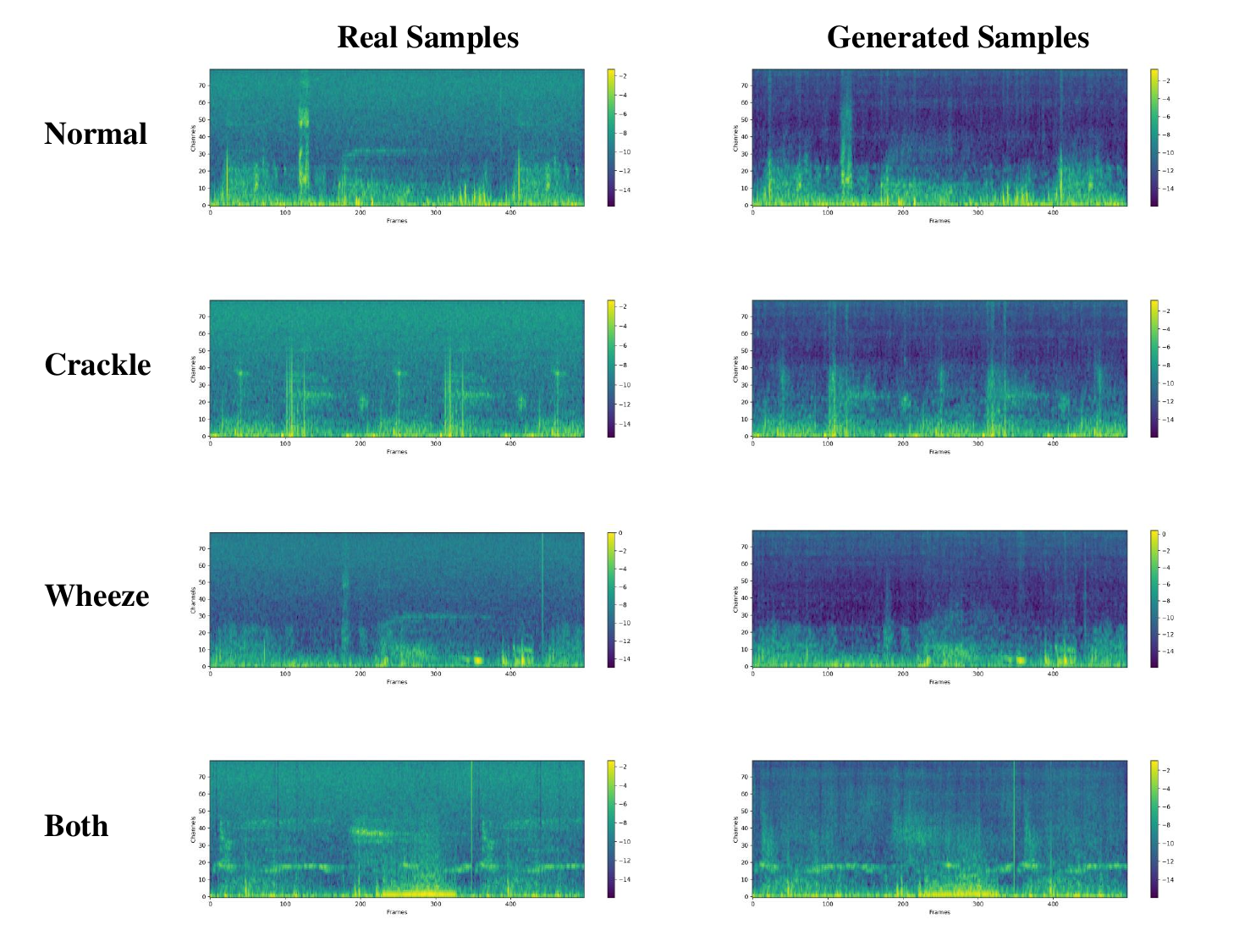} 
    \caption{Comparison of spectrograms per each class randomly chosen from the test set and the generated results.}
    \label{fig:fig2}
\end{figure}

Figure~\ref{fig:fig2} provides visual comparison of spectrograms randomly sampled per class from the test set and the results generated by our diffusion model when conditioned on these spectrograms. The generated spectrograms per class are visually similar to the original sample, which demonstrates the capability to generate high-fidelity audio, yet introduce small realistic variations that provide some value for augmentation.

\section{Conclusion}
We presented a simple method for generating realistic respiratory sound samples using an audio diffusion model. 
We further introduced adversarial fine-tuning to address the distribution inconsistency between synthetic and real samples. 
Our results show that our method can effectively improve the performance of imbalanced abnormal classes, demonstrating its ability to address the challenges of using synthetic data.
%Our experiments show that the proposed adversarial fine-tuning method successfully aligns features between synthetic and original respiratory sound samples by training a discriminator to distinguish between the two types of samples, while simultaneously training a classifier to predict the respiratory sound label.
%Furthermore, our results demonstrated that the proposed adversarial fine-tuning method enables more effective learning of respiratory sound classification models, even when the generated samples used for data augmentation have a different data distribution from the real samples. --> I think this sentence is little dangerous, right?
We believe that our method can be helpful in various other datasets and could be used to supplement other augmentation methods.
%We believe that our method can be employed in various medical datasets.

%\begin{itemize}
%\end{itemize}

\begin{ack}
This research was supported by the MSIT(Ministry of Science and ICT), Korea, under the ITRC(Information Technology Research Center) support program(IITP-2023-2020-0-01808) supervised by the IITP(Institute of Information \& Communications Technology Planning \& Evaluation), and by Brian Impact, a non-profit organization dedicated to the advancement of science and technology.

\end{ack}

\bibliography{references}
\bibliographystyle{plain}

%%%%%%%%%%%%%%%%%%%%%%%%%%%%%%%%%%%%%%%%%%%%%%%%%%%%%%%%%%%%
\vfill\pagebreak
\appendix

\section{Related Works}
\label{appendix_related}
\paragraph{Respiratory Sound Classification} The ICBHI~\cite{rocha2018alpha} dataset is a well-known benchmark for respiratory sound classification. Various neural network-based approaches have been developed for this task, including residual blocks~\cite{gairola2021respirenet, nguyen2022lung, wang2022domain}, CNN~\cite{moummad2022supervised}, pretrained models on ImageNet~\cite{gairola2021respirenet, nguyen2022lung, wang2022domain}, AudioSet~\cite{moummad2022supervised}, and Audio Spectrogram Transformer (AST)~\cite{gong21b_interspeech}. To address the challenge of limited data, previous studies have proposed various learning protocols, including device-specific fine-tuning~\cite{gairola2021respirenet}, mixup as well as splicing audio augmentation~\cite{wang2022domain}, task-specific co-tuning~\cite{nguyen2022lung}, supervised contrastive learning~\cite{moummad2022supervised}, and patch-mix contrastive learning~\cite{bae23b_interspeech}. Instead of focusing on previous data augmentation methods, this paper addressed the challenge of using synthetic samples generated by deep generative models. To this end, we first trained a pre-trained AST~\cite{bae23b_interspeech} model on the ICBHI dataset as described in~\cite{gong21b_interspeech}. We also trained the model on the Mixed-ICBHI dataset, which contains both synthetic and real samples. We then showed that our proposed adversarial fine-tuning method can overcome the data distribution inconsistency between synthetic and real samples.
%In this paper, we first evaluated the performance of pretrained AST~\cite{gong21b_interspeech} on the ICBHI dataset as described in~\cite{bae23b_interspeech} using synthetic samples. %We then showed the proposed adversarial fine-tuning can overcome the data distribution mismatch between synthetic and real samples. 

\paragraph{Deep Generative Models} Recent advances in DGMs, such as GAN~\cite{goodfellow2014generative}, VAE~\cite{kingma2013auto}, and diffusion models~\cite{ho2020denoising}, have attracted significant attention. This is because DGMs can be used to generate synthetic samples to mitigate data scarcity issues. They have been applied to medical images, such as retinal images~\cite{costa2017end,iqbal2018generative}, CT and MRI scans~\cite{nie2017medical, shin2018medical, bowles2018gan, sandfort2019data, han2019synthesizing, armanious2020medgan, wolleb2022diffusion, dorjsembe2022threedimensional}, and X-rays~\cite{loey2020within, motamed2021data} which received additional interest due their applicability in diagnosing COVID-19 cases. Several approaches have also been introduced to generate synthetic sequential medical data, such as respiratory sounds~\cite{kochetov2020generative, jayalakshmy2021conditional, saldanha2022data}, EEG recordings~\cite{fahimi2019towards, hazra2020synsiggan, tian2023dual}, and ECG signals~\cite{narvaez2020synthesis, wulan2020generating}.
Unlike previous studies on respiratory sound, our work was the first attempt to successfully generate high-fidelity respiratory sound samples using an audio diffusion model~\cite{kong2020diffwave} which is neural vocoding conditioned on Mel-spectrogram as a conditional neural vocoder.

\section{ICBHI Dataset Details}
\label{appendix_icbhi}
\begin{table}[!ht]
    \centering
    \caption{Overall details of the ICBHI~\cite{rocha2018alpha} respiratory sound dataset.}
    % \vspace{-1pt}
    \label{app_tab:tab1}
    \addtolength{\tabcolsep}{10pt}
    \footnotesize{
    %\begin{tabular}{p{1pt}lccc}
    \begin{tabular}{l|cc|c}
    \toprule
    & \multicolumn{3}{c}{number of respiratory samples (ratio)} \\
    \cmidrule(l{2pt}r{2pt}){2-4}
    label & train & test & sum \\
    \hline 
    \midrule
    Normal & 2,063 (49.8\%) & 1,579 (57.29\%) & 3,642 \\
    Crackle & 1,215 (29.3\%) & 649 (23.55\%) & 1,864 \\
    Wheeze & 501 (12.1\%) & 385 (13.97\%) & 886 \\
    Both & 363 (8.8\%) & 143 (5.19\%) & 506 \\
    \midrule
    Total & 4,142 & 2,756 & 6,898 \\
    \bottomrule
    \end{tabular}}
\end{table}
\begin{table}[!ht]
    \centering
    \caption{Overall details of Mixed-ICBHI dataset with synthetic and real samples.}
    % \vspace{-1pt}
    \label{app_tab:tab2}
    \addtolength{\tabcolsep}{10pt}
    \resizebox{\linewidth}{!}{
    %\begin{tabular}{p{1pt}lccc}
    \begin{tabular}{l|ccccccc}
    \toprule
    & \multicolumn{7}{c}{mixed dataset (synthetic ratio, \%)} \\
    \cmidrule(l{2pt}r{2pt}){2-8}
    label   & Mixed-500     & Mixed-800     & Mixed-1k     & Mixed-1.5k     & Mixed-2k       & Mixed-3k       & Mixed-5k       \\
    \hline 
    \midrule    
    normal  & 0\      & 0\      & 0 0      & 0        & 0        & 31.23  & 58.74 \\
    crackle & 0\      & 0\      & 0     & 19.00     & 41.11  & 59.50 & 75.70 \\
    wheeze  & 0       & 37.38 & 49.90 & 66.60  & 75.72 & 83.30 & 89.98 \\
    both    & 27.40 & 54.63 & 63.70 & 75.80 & 82.40 & 87.90 & 92.74 \\
    \bottomrule
    \end{tabular}}
\end{table}

The ICBHI~\cite{rocha2018alpha} dataset is a well-known benchmark for respiratory sound classification. The ICBHI dataset consists of 6,898 respiratory cycles, with a total duration of approximately 5.5 hours. The dataset is officially split into a training set (60\%) and a test set (40\%), with no patient overlap between the two sets. As shown in Table~\ref{app_tab:tab1}, the training and test sets contain 4,142 and 2,756 samples respectively and are categorized into four classes, \emph{normal} (49.8\%/57.29\%), \emph{crackle} (29.3\%/23.55\%), \emph{wheeze} (12.1\%/13.97\%) and \emph{both} (8.8\%/5.19\%), respectively. For all our experiments, we resampled all the samples to 16 kHz. For the metrics, we used \textit{Sensitivity} ($S_{e}$), \textit{Specificity} $(S_{p})$, and their arithmetic mean \textit{Score} as described in~\cite{rocha2018alpha}.

\section{Mixed Dataset Details}
\label{appendix_mixed_dataset}
%To address the limitations of the training data, which consisted of a small number of samples with imbalanced class distributions, we generated synthetic samples to achieve a uniform class distribution of 500 to 5,000 samples per class. 
As described in Table~\ref{app_tab:tab2}, we mixed the synthetic samples with the real data to create Mixed-ICBHI datasets as follows: \emph{Mixed-500, ..., Mixed-N, ..., Mixed-2k} where the number $N$ refers to the total amount of samples per class. We prioritize real samples so that synthetic samples are only added if the sample count is less than $N$ (i.e., \emph{Mixed-500} only contains synthetic samples from $C_{both}$).

\section{Confusion Matrices Results}
\label{appendix_confusion}
\begin{figure}[!h]
    \centering
    \begin{subfigure}{.5\linewidth}
      \centering
      \includegraphics[width=1.0\linewidth]{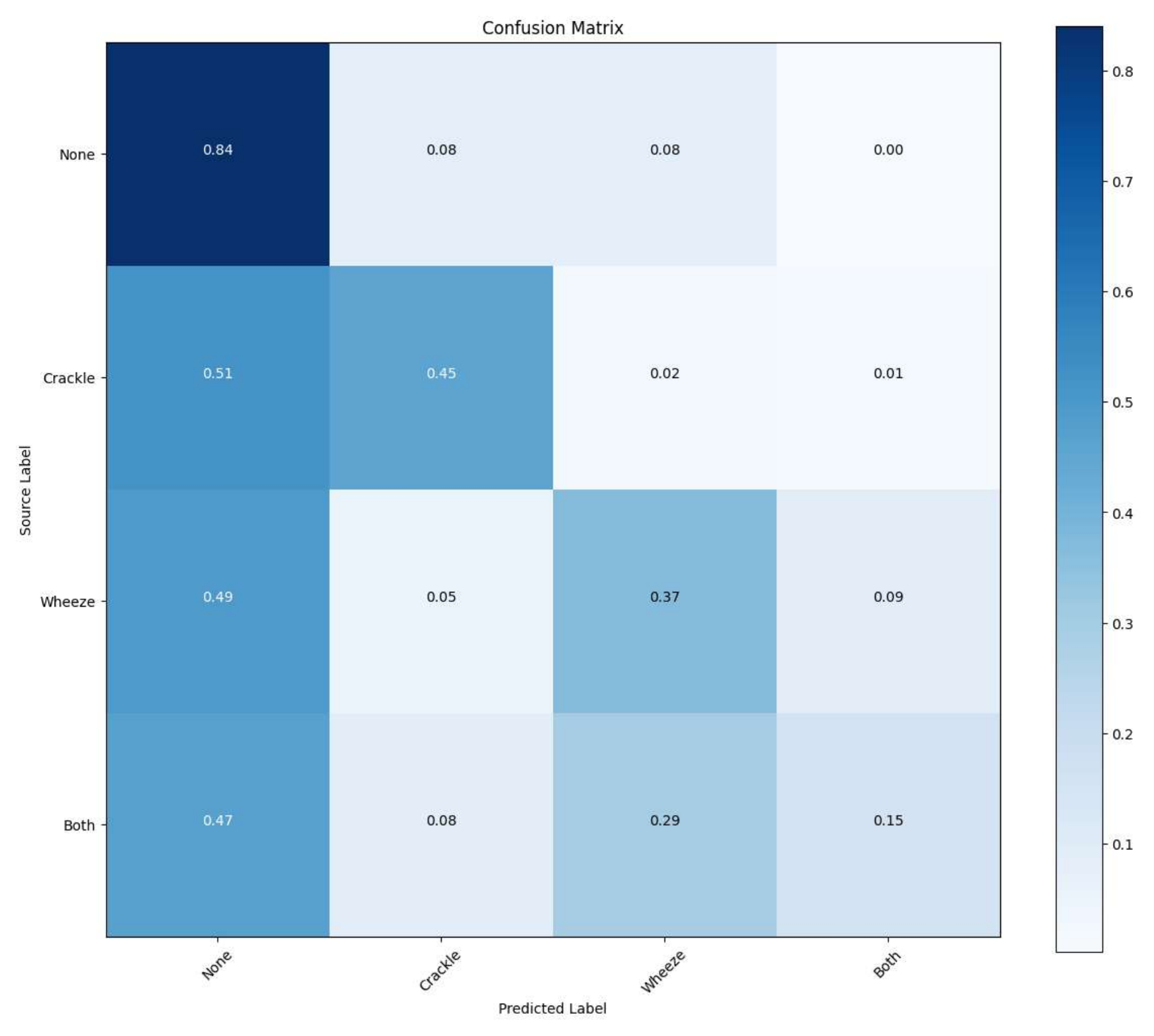}
      \caption{AST FT (No Aug.)}
      \label{fig:sfig1}
    \end{subfigure}%
    \begin{subfigure}{.5\linewidth}
      \centering
      \includegraphics[width=1.0\linewidth]{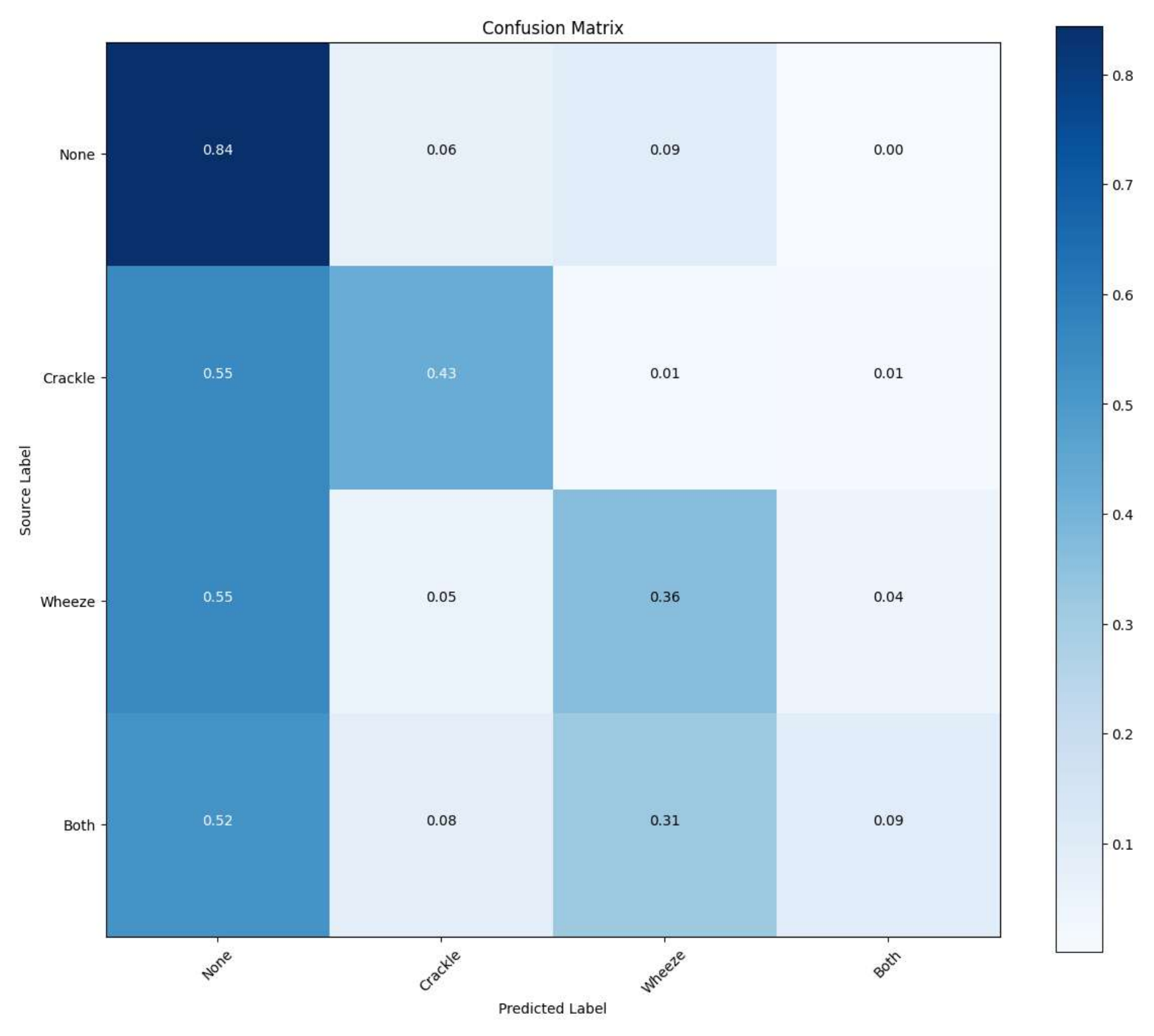}
      \caption{AST FT (Mixed-500)}
      \label{fig:sfig2}
    \end{subfigure}
    \begin{subfigure}{.5\linewidth}
      \centering
      \includegraphics[width=1.0\linewidth]{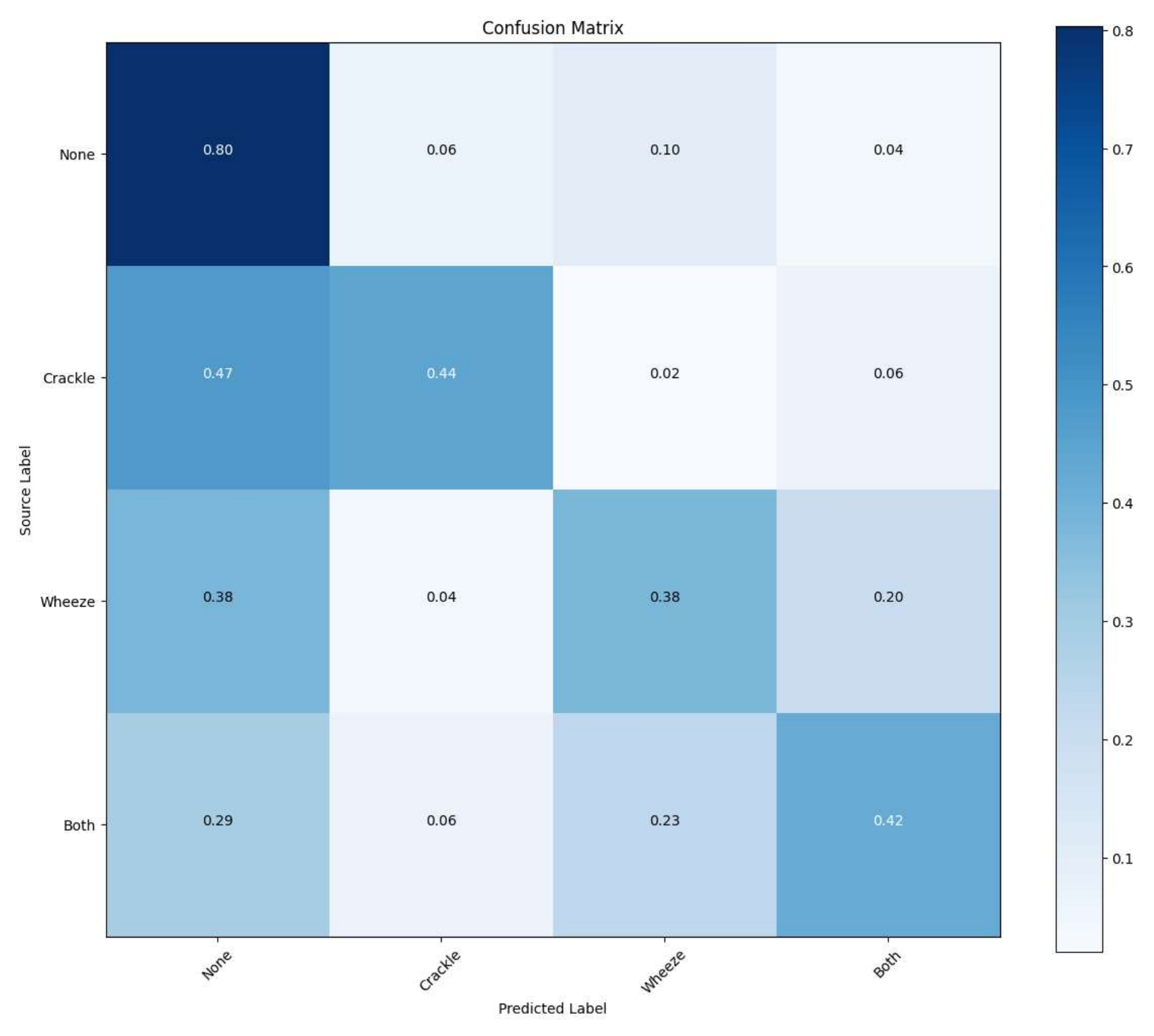}
      \caption{AFT (Mixed-500)}
      \label{fig:sfig3}
    \end{subfigure}
    \caption{Confusion matrix results of AST FT with no augmentation, AST FT and AFT with Mixed-500 and Mixed-2k, respectively.}
    \label{fig:confusion_matrices}
    
    \end{figure}

To show how the proposed method affects \emph{all the classes}, Figure~\ref{fig:confusion_matrices} provides the confusion matrices between the AST FT with no augmentation, AST FT and AFT with Mixed-500 and Mixed-2k, respectively. The proposed method did not degrade considerably on normal classes and achieved the highest performance compared to other methods on the most imbalanced classes. Our results demonstrate the effectiveness and potential of our proposed method, showing its ability to address the data distribution inconsistency problem with synthetic data, especially in class imbalanced problems.

%We argue that the proposed adversarial adaptation method enables effective learning for respiratory sound classification, even though the generated samples used for data augmentation may have a different data distribution from the real samples.

\vfill\pagebreak

\end{document}